\documentclass[twocolumn,secnumarabic,amssymb, nobibnotes, aps, prl]{revtex4-2}

\usepackage[dvipdfmx]{graphicx}
\usepackage{color}
\usepackage{amsmath}

\begin{document}

\title{Interplay of thermal and non-thermal effects  in x-ray-induced ultrafast melting}%

\author{Ichiro Inoue$^{1}$}%
\email{inoue@spring8.or.jp}
\author{Victor Tkachenko$^{2,3}$}
\email{victor.tkachenko@xfel.eu}
\author{Yuya Kubota$^{1}$}
\author{Fabien Dorchies$^{4}$}
\author{Toru Hara$^{1}$}
\author{Hauke H\"{o}eppner$^{5}$}
\author{Yuichi Inubushi$^{1, 6}$}
\author{Konrad J. Kapcia$^{3,7}$}
\author{Hae Ja Lee$^{8}$}
\author{Vladimir Lipp$^{9,3}$}
\author{Paloma Martinez$^{4}$}
\author{Eiji Nishibori$^{10,11}$}
\author{Taito Osaka$^{1}$}
\author{Sven Toleikis$^{12}$}
\author{Jumpei Yamada$^{1}$}
\author{Makina Yabashi$^{1,6}$}
\author{Beata Ziaja$^{3,9}$}
\email{beata.ziaja-motyka@cfel.de}
\author{Philip A. Heimann$^{8}$}
\email{paheim@slac.stanford.edu}

\affiliation{$^1$RIKEN SPring-8 Center, 1-1-1 Kouto, Sayo, Hyogo 679-5148, Japan\\
$^2$European XFEL GmbH, Holzkoppel 4, 22869 Schenefeld, Germany\\
$^3$Center for Free-Electron Laser Science CFEL, Deutsches Elektronen-Synchrotron, Notkestr.  85, 22607 Hamburg, Germany\\
$^4$Universit\'{e} de Bordeaux, CNRS, CEA, Centre Lasers Intenses et Applications, UMR 5107, F-33400 Talence, France\\
$^5$Helmholtz-Zentrum Dresden-Rossendorf, 01328 Dresden, Germany\\
$^6$Japan Synchrotron Radiation Research Institute, Kouto 1-1-1, Sayo, Hyogo 679-5198, Japan\\
$^7$Institute of Spintronics and Quantum Information, Faculty of Physics, Adam Mickiewicz University in Pozna$\acute{n}$, Uniwersytetu Pozna$\acute{n}$skiego 2, PL-61614 Pozna$\acute{n}$, Poland\\
$^8$SLAC National Accelerator Laboratory, Menlo Park, California 94025, USA\\
$^9$Institute of Nuclear Physics, Polish Academy of Sciences, Radzikowskiego 152, 31-342 Krakow, Poland\\
$^{10}$Graduate School of Pure and Applied Sciences, University of Tsukuba, Tsukuba, Ibaraki 305-8571, Japan\\
$^{11}$Faculty of Pure and Applied Sciences and Tsukuba Research Center for Energy Materials Science, University of Tsukuba, Tsukuba, Ibaraki 305-8571, Japan\\
$^{12}$Deutsches Elektronen-Synchrotron, Notkestr.  85, 22607 Hamburg, Germany}

\begin{abstract}
X-ray laser-induced structural changes in silicon undergoing femtosecond melting have been investigated by using an x-ray pump-x-ray probe technique. The experimental results for  different initial sample temperatures reveal that the onset time and the speed of the atomic disordering are independent of the initial temperature, suggesting that equilibrium atomic motion in the initial state does not play a pivotal role in the x-ray-induced ultrafast melting. By comparing the observed time-dependence of the atomic disordering and the dedicated theoretical simulations, we interpret that the energy transfer from the excited electrons to ions via electron-ion coupling (thermal effect) as well as 
a strong modification of the interatomic potential  due to electron excitations (non-thermal effect) trigger the ultrafast atomic disordering.
Our finding of the  interplay of thermal and non-thermal effects in the x-ray-induced melting  demonstrates that accurate modeling of intense x-ray interactions with matter is essential to ensure a correct interpretation of experiments using intense x-ray laser pulses.
\end{abstract}

\maketitle

Ultrafast laser pulses can bring matter into highly non-equilibrium states and induce exotic processes. The well-known example is the femtosecond melting, which is  often called non-thermal melting.
 It has been observed in various semiconductors \cite{Shank1983, Shank1983_2, Glezer1995, Siders1999, Rousse2001, Sokolowski2001, Lindenberg2005, Gaffney2005, Hillyard2007, Harb2008, Wang2020, Bengtsson2020} and in two-dimensional materials \cite{Vorobeva2011,Porer2014} irradiated with femtosecond optical laser pulses, where the excitation of a large fraction (more than a few percent) of the valence electrons modifies the interatomic potential and drives ultrafast atomic disordering without equilibrium between the electron and ion subsystem \cite{Sundaram2002}.
 
The recent advent of x-ray free-electron lasers (XFELs)\cite{McNeilNP2010, Pellegrini2016}, emitting intense femtosecond x-ray pulses, has extended these studies to the x-ray regime. Understanding the physics governing the x-ray-induced ultrafast melting is of great importance in the context of practical applications of XFELs, particularly the structure determination of  nanocrystals  \cite{Chapman2011,Ilme2015}. While x-ray nanocrystallography of organic molecules and proteins is challenging with conventional light sources, because radiation damage perturbs their structure before completing the data collection \cite{Owen2006, Howells2009, Garman2010, Holton2010}, the short duration of the XFEL pulses allows one to measure the diffraction signal before the manifestation of radiation damage \cite{Chapman2014}. In those experiments, the XFEL pulses are focused down to a micrometer size (or even less) so that the beam size matches the crystal size. Since the irradiation with the focused XFEL pulses inevitably excites many valence electrons to the conduction band \cite{Medvedev2018, Inoue2021, Victor2021}, setting the pulse duration to be shorter than the onset time of the atomic disordering is essential for the success of such experiments.

The key question that remains controversial is which factors determine the speed of atomic disordering in x-ray-excited materials. 
For example, the x-ray-induced femtosecond melting in silicon (Si) has been intensively studied both experimentally \cite {Pardini2018, Hartley2021} and theoretically \cite{Medvedev2015, Victor2021}, but the detailed mechanism of the disordering processes is still under debate.
An x-ray pump-x-ray probe experiment \cite {Pardini2018} showed that the root-mean-square (rms) atomic displacements in x-ray-excited Si increase with time at a constant rate nearly equal to the velocity of atoms in the equilibrium state ($\sqrt{3k_BT/m}$, where $k_B$ is the Boltzmann constant, $T$ is the sample temperature before x-ray excitation, $m$ is the mass of atom).
From the observed results, the authors of \cite{Pardini2018} claimed the inertial atomic motion and concluded that the thermal atomic motion in the equilibrium state determines the speed of atomic disordering. In another experiment,  Hartley \textit{et al.} \cite{Hartley2021} measured the time-dependence of the diffuse scattering in Si after irradiation with an XFEL pulse. By comparing the experimental results and simulations, they concluded that the modified interatomic potential due to the electron excitation (``non-thermal effect") dominates the speed of the atomic disordering.
Recent numerical simulations that investigated the damage threshold of x-ray-induced ultrafast melting in various materials \cite{Medvedev2015, Medvedev2019, Medvedev2020} suggested that not only the non-thermal effect but also the energy transfer from the excited electrons to ions via electron-ion coupling (``thermal effect") may contribute to the atomic disordering on the femtosecond timescale.

\begin{figure}
\includegraphics[width=8.5cm]{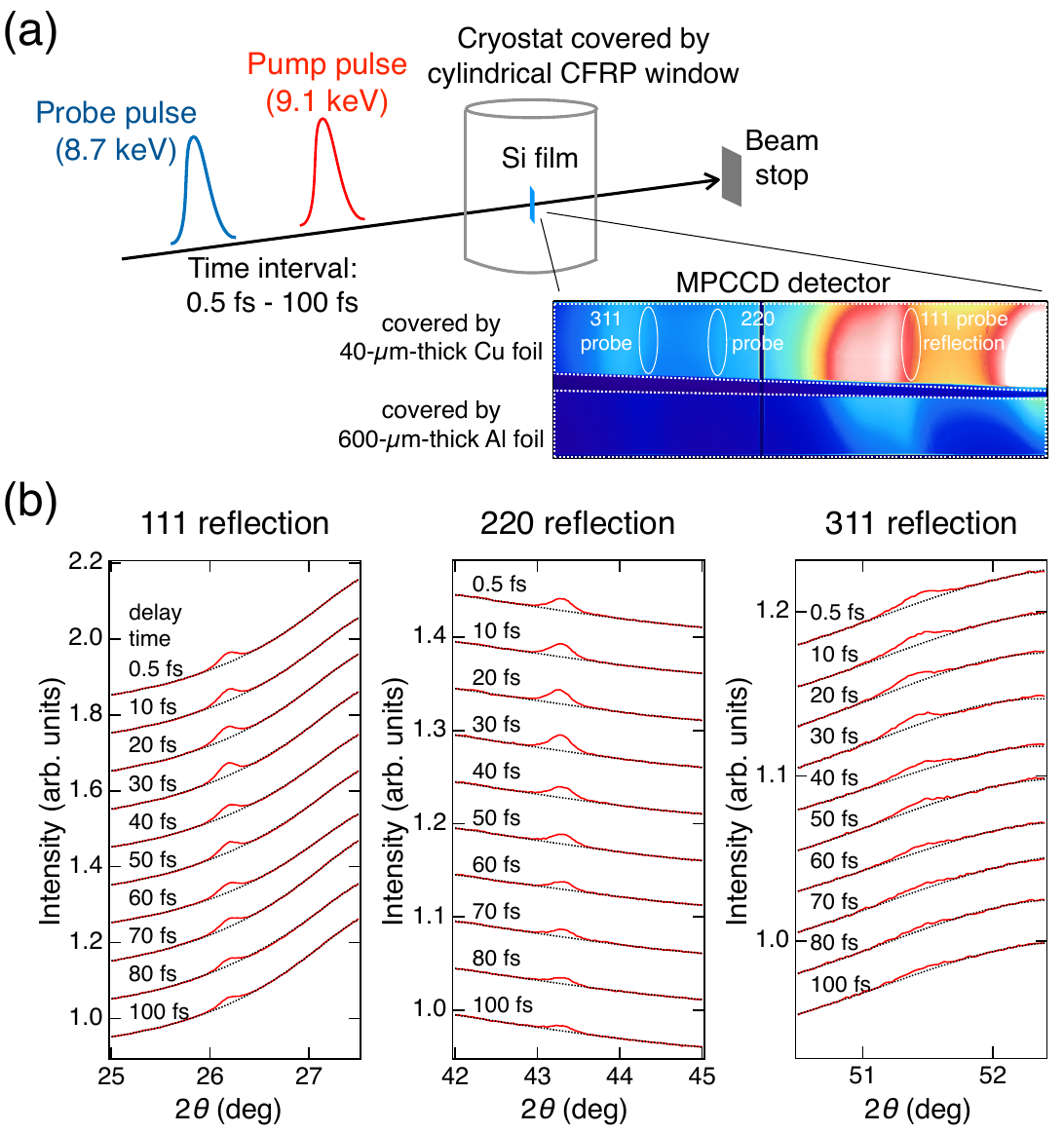}
\caption{(a) A schematic illustration of the experiment.
(b) The diffraction profiles of probe pulses for silicon at 300 K. For better visibility, each diffraction profile is linearly shifted along the vertical axis. Black dotted lines show the estimated background.}
\end{figure}

In order to elucidate the impact of the three aforementioned factors (thermal effect, non-thermal effect, and initial equilibrium atomic motion) on the speed of atomic disordering, we conducted measurements of XFEL-induced structural changes in silicon at various temperatures
 by using an x-ray pump-x-ray probe technique \cite{Inoue2016, Inoue2021, Inoue2022}. By exploiting ultrafast XFEL pulses from SACLA with duration of much below 10 fs  \cite{IshikawaNP2012, Inubushi2017, InouePRAB2018, InoueJSR2019, Osaka2022},
we captured the initial stage of the atomic disordering. From the comparison between the experimental results and dedicated simulations, we can understand the physics governing x-ray-induced ultrafast melting.

Figure 1 (a) shows a schematic illustration of the experimental setup at SACLA BL3 \cite{YabashiJSR2015}. The XFEL machine was operated to generate 9.10-keV pump and 8.70-keV probe pulses with the rms duration of 2.5 fs by a split-undulator scheme \cite{Hara2013}. Since the jitter of the time interval between the double pulse  was negligibly small (much less than 1 fs \cite{Hara2013}), the x-ray pulse duration determined the time resolution of the measurement ($\sqrt{2}\times$ 2.5 fs = 3.6 fs). The pump and probe pulses were focused to FWHM sizes of 1.8 $\mu$m (horizontal) $\times$ 1.8 $\mu$m (vertical) and 1.0 $\mu$m (horizontal) $\times$ 1.4 $\mu$m (vertical), respectively, by using an x-ray mirror system \cite{Yumoto2013, Tono2017}. We used a 10-$\mu$m-thick nanocrystal Si film (grain size of 500 nm, US research nanomaterials) as the target. The Si film attached to a polyimide film was set to a helium closed-cycle cryostat with a cylindrical x-ray window made of carbon fiber reinforced plastics (CFRP). The sample was placed at the focus and continuously translated spatially so that each double pulse irradiated the undamaged surface. The diffraction peaks from the sample (111, 220, and 311 reflections) in the horizontal plane were measured by using a multiport charge-coupled device (MPCCD) detector \cite{KameshimaRSI2014} that covered  the scattering angle (2$\theta$) range of 20$^\circ$-55$^\circ$. The top-half of the detector was covered by a 40-$\mu$m-thick copper foil such that the diffraction signals from the probe pulses selectively impinged on the detector, 
while  a 600-$\mu$m-thick aluminum foil covered bottom-half of the detector to measured the diffraction signals from the pump pulses.
The shot-by-shot pulse energy at the sample was characterized by using an inline spectrometer \cite{Tamasaku2016} located upstream of the focusing mirror system and taking into account the reflectivity of the mirrors and the transmittance of the CFPR window. The diffraction data at initial sample temperature $T=$ 10, 100 and 300 K were collected by changing the delay time from 0.5 fs to 100 fs.

We extracted the detector data for double pulses with specific pulse energies of the pump (100$\pm$20 $\mu$J) and probe pulses (30$\pm$20 $\mu$J) and calculated the averaged diffraction profile for each temperature and delay time.
The absorbed dose for the selected pump pulses was $\sim$10 eV/atom, which was higher than the predicted damage threshold of Si ($\sim$1 eV/atom) \cite{Medvedev2015}.
Figure 1 (b) shows the probe diffraction profiles in the vicinity of each reflection peak for the experiment at $T=$300 K. 
It is clearly seen  that the diffraction intensity decreased with the delay time for each reflection index.
We can consider two possible reasons for the ultrafast decay of the diffraction intensity (the change in atomic scattering factor and the progressing atomic 
disorder). One is the change in atomic scattering factors caused by the photoionization and secondary ionizations \cite{Ziaja2001, Ziaja2005, Son2011, Ziaja2015, Inoue2023}. 
A theoretical calculation showed that the atomic scattering factors of the ionized atoms are mostly determined by the number of occupied core levels \cite{Hau-Riege2007}. 
However, the number of core holes per atom is too small ($\sim10^{-2}$ for the x-ray dose in the current experiment \cite{Victor2021})  to explain the experimental observations.
The only possible origin for the diffraction intensity decay is the progressing atomic disorder. 
Since the Bragg diffraction angle for the probe pulses did not change for all delay times, it is natural to consider that the lattice constants  did not change on the femtosecond time scale.
In this case, the diffraction intensity may be proportional to $\exp (-q^2 \langle u_{hkl}^2\rangle)$ with  the mean square of the atomic displacement perpendicular to the ($hkl$) plane  $\langle u_{hkl}^2\rangle$ and the scattering vector $q=4\pi\sin\theta/\lambda$ with the wavelength $\lambda$, as an analogy to the Debye-Waller factor in crystallography.

\begin{figure}
\includegraphics[width=5.5cm]{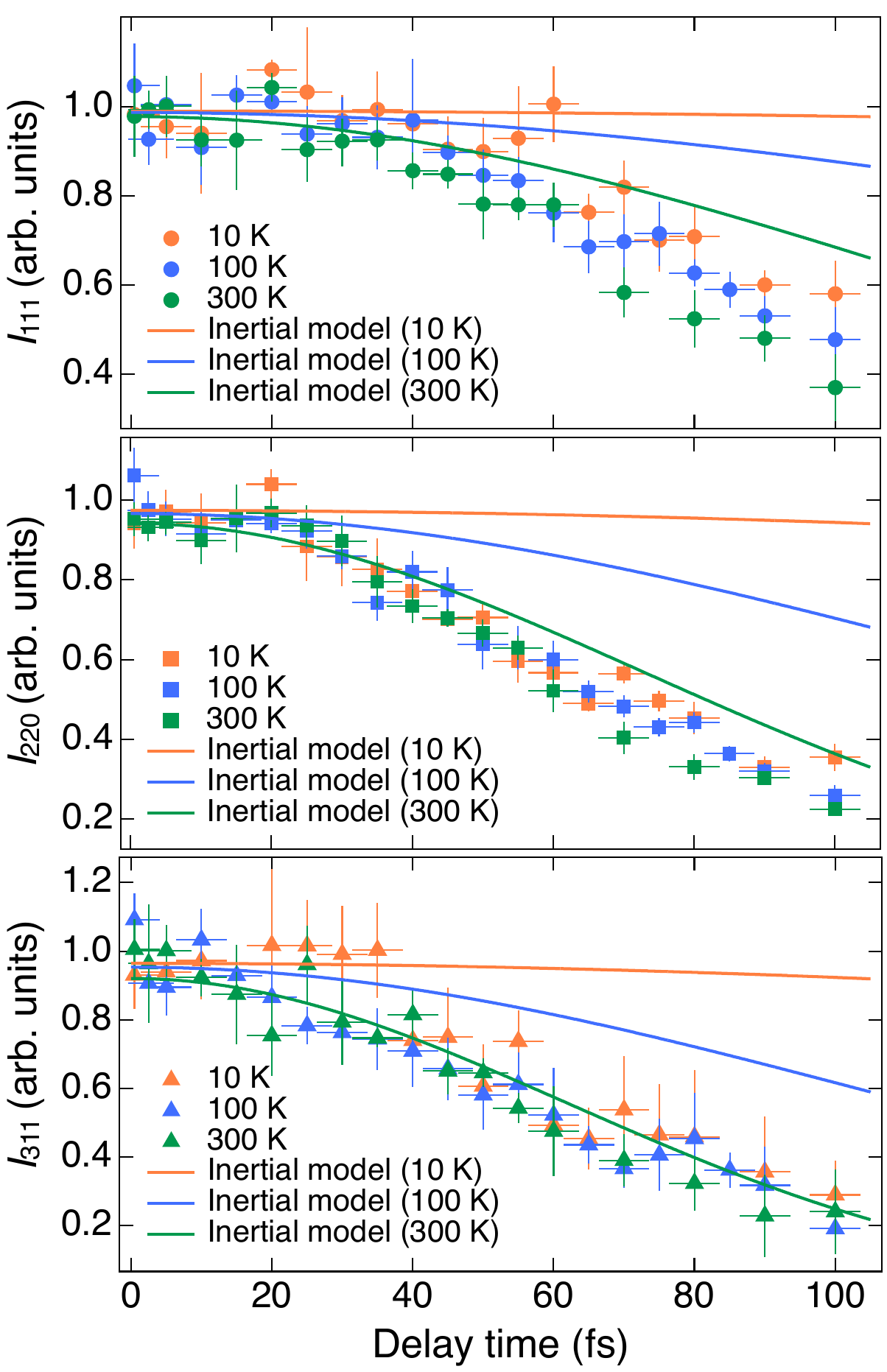}
\caption{Probe diffraction efficiency as a function of delay time for different sample temperatures (orange markers: 10 K, blue markers: 100 K, and green markers: 300K). Solid curves represent the diffraction efficiency predicted by the inertial mode.}
\end{figure}

We evaluated the atomic disorder from the measured diffraction profiles of the pump and probe pulses as follows. First, the background of the diffraction profiles of the probe pulses was estimated by fitting the profiles in the vicinity of diffraction peaks with polynomial functions (black dotted curves in Fig. 1(b)). After subtracting the estimated background,  each diffraction peak was fitted by a Gaussian function and the integrated diffraction intensity ($I^{probe}_{111}$, $I^{probe}_{220}$, $I^{probe}_{311}$) was determined. For comparing the probe diffraction intensity between different delay times, we compensated  the inhomogeneity of the sample thickness and the differences in the probe pulse energy by calculating the diffraction efficiency of the probe pulses given by $I_{hkl}=\frac{I^{probe}_{hkl}/E^{probe}}{I^{pump}_{220}/E^{pump}}$, where $hkl$ represents the reflection index,  $E_{pump(probe)}$ is the average pulse energy of the pump (probe) pulses on the sample, and $I^{pump}_{220}$ is the pump diffraction intensity of the 220 reflection evaluated by the same procedures for the pump diffraction intensity. Here, we used $I^{pump}_{220}$, rather than the pump diffraction intensity of other reflections because the uncertainty of $I^{pump}_{220}$ was smaller than those for other reflections. 
Figure 2 shows the probe diffraction efficiency as a function of delay time.
Here, the diffraction efficiency for each initial sample temperature and for each reflection is normalized such that the averaged value for the short delay times (0.5-15 fs) equals to the Debye-Waller factor  $\exp(-q^2\langle u_0^2\rangle)$ for the undamaged Si calculated with
$\langle u_0^2\rangle=$
$2.4\times 10^{-3}$  \AA$^2$ (10 K) \cite{Reid1980}, $3.2\times 10^{-3}$ \AA$^2$ (100 K) \cite{Wahlberg2016}, $5.6\times 10^{-3}$ \AA$^2$ (300 K) \cite{Wahlberg2016}.
The vertical error bars represent the standard deviation of the diffraction efficiency calculated for five independent sub-ensembles from the whole extracted pulses. The diffraction efficiency for all temperatures was almost the same at each delay time, indicating that thermal atomic motion in the equilibrium state is not related to the speed of the atomic disordering. Negligible contribution of the thermal atomic motion to the ultrafast melting can be also confirmed by comparing the experimental results and the diffraction efficiency predicted by the inertial model \cite{Pardini2018} given by
\begin{equation}
I_{hkl}=
\begin{cases}
\exp(-q^2\langle u_0^2\rangle) & \text{if  $t\le t_0$}\\
\exp(-q^2 (\langle u_0^2\rangle+v^2(t-t_{0})^2)) & \text{otherwise},
\end{cases}
\end{equation}
where $t_0$ is the onset time of the atomic displacement and $v=\sqrt{k_BT/m}$=5.4$\times 10^{-4}$ (10 K), 1.7 $\times 10^{-3}$ (100 K), and 3.0 $\times 10^{-3}$ \AA/fs (300 K). Even if we select the onset to be $t_0=0$ fs, the diffraction efficiency predicted by the inertial model is higher than that observed in the experiment (Fig. 2).

\begin{figure}
\includegraphics[width=8.5cm]{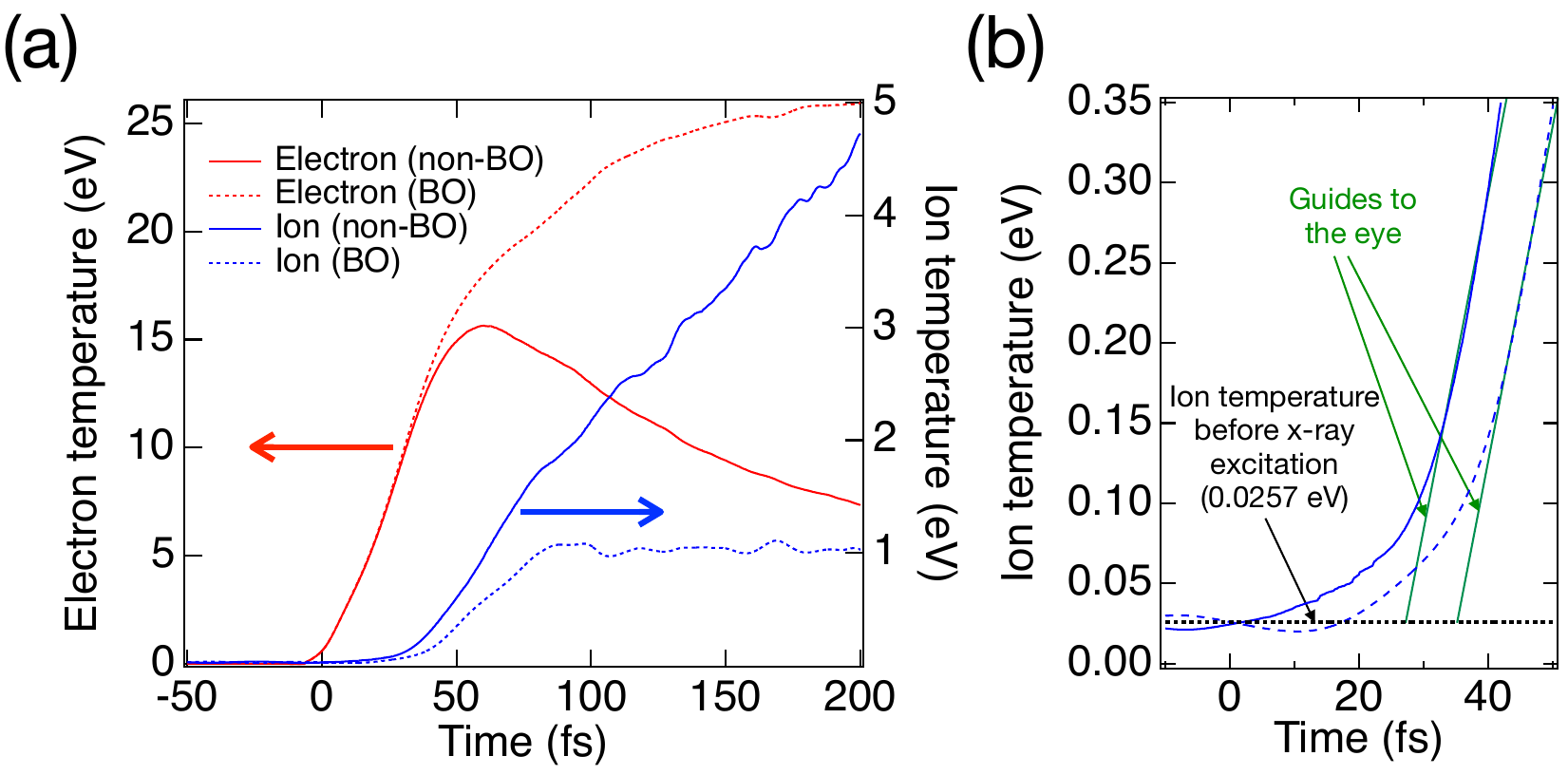}
\caption{(a) Simulated ion and electron temperatures in x-ray-excited silicon of initial temperature of 300 K.  Time zero in the horizontal axis corresponds to the intensity maximum of the pump pulse. (b) Enlarged plot of the ion temperatures around time zero.}
\end{figure}

Next, we discuss the thermal and non-thermal effects in the ultrafast melting  by comparing the experimental results and theoretical simulations performed by using \textit{XTANT} (x-ray-induced thermal and non-thermal transitions) code \cite{Medvedev2013_2, Medvedev2018_2}. First, we simulated the electron and ion temperatures of the x-ray-excited Si within and without Born-Oppenheimer approximation, which excludes or includes electron-ion coupling respectively (BO- and non-BO-simulations). The calculation was performed for a 216-atom-large supercell at 300 K irradiated with spatially uniform x-ray pulse with 6 fs duration (FWHM) and fluence corresponding to the average fluence in the experiment (pulse energy divided by the product of horizontal and vertical FWHM beam sizes).
 For both simulations, the ions keep the original temperature for the first few tens of fs after irradiation with the pump pulse. Then, the ion temperature quickly increases well above the original sample temperature (Fig. 3 (a)), which is consistent with our experimental observation that the equilibrium atomic motion did not significantly contribute to the speed of atomic disordering. One major difference in the results for the two simulations is the onset time of the increase in ion temperature (Fig. 3(b)); the ion temperature starts to rapidly increase at $~25$ fs after the x-ray exposure in the non-BO simulation, while the onset time of such temperature increase is $~35$ fs in the BO simulation. This result confirms that both the thermal effect as well as the non-thermal effect contribute to the progressing atomic disordering, i.e., the interplay between the thermal and non-thermal effects is present in the x-ray-induced ultrafast melting.

The contribution of thermal effects to ultrafast melting can be experimentally confirmed by comparing the simulated and experimentally measured onset times of atomic disordering. Figures 4 (a)-(c) show the rms atomic displacements perpendicular to the ($hkl$) plane ($hkl$=111, 220, 311) in the BO- and the non-BO simulations of Si at 300 K performed with the fluences of (a) 100\%, (b) 48\%, and (c) 16\% of the average fluence in the experiment. For all fluence conditions and reflection indices, the onset time of atomic disordering predicted by the non-BO simulation is 25-30 fs, which is approximately 10 fs faster than the predicted value in the BO simulation. Since the difference in the onset time is larger than the time resolution in the present experiment, our experimental data can be readily used to check the existence of thermal effects in ultrafast melting. Figure 4 (d) shows the experimentally observed rms atomic displacement perpendicular to the (220) plane ($\sqrt{\langle u_{220}^2 \rangle}$) evaluated through the relationship $I_{220}=\exp(-q^2\langle u_{220}^2 \rangle)$.  
We here evaluated the atomic displacement from the observed diffraction efficiency of only 220 reflection, because it was measured with higher accuracy compared to other reflections (Fig. 2).
It is clearly seen that the experimentally observed atomic disorder starts to increase at  $\sim$25 fs after irradiation with the pump pulse, irrelevant of the initial sample temperature. The onset time of atomic disordering evaluated by the experiment is consistent with the prediction of  the non-BO simulation rather than the BO simulation, supporting that both thermal and nonthermal effects contribute to the initial processes in x-ray-induced ultrafast melting.

Although the non-BO simulation reproduces the onset time of atomic disordering as evaluated by the experiment, the predicted degree of atomic disorder for the average fluence (Fig. 4(a)) is higher than the experimentally measured values (Fig. 4(d)). This discrepancy may be explained by the non-uniformity of the pump fluence. Since the focal spots of the pump and probe pulses had Gaussian shapes in the present experiment \cite{Yumoto2013}, the probe diffraction signals originated from various sample volumes irradiated with different pump fluence. The simulation results shown in Figs. 4(a)-(c) predict that the atomic disordering of crystals experiencing higher pump fluence increases faster, implying that their contribution to the probe diffraction intensity decreases with time. Consequently, the ``effective" pump fluence might be reduced with the increasing delay time in the present experiment.  In fact, we can see that the results of non-BO simulation obtained with the fluence corresponding to the 48\% of the effective fluence (Fig. 4 (b)) are close to the experimental results (Fig. 4(d)). 
Let us emphasize that summing up the scattering amplitude over the entire sample volume irradiated with the probe pulse, as done in our previous study \cite{Victor2021}, would require an extensive and long-taking computational effort, not bringing new aspects for the data understanding. We leave it therefore out in the present paper.

\begin{figure}
\includegraphics[width=8.5cm]{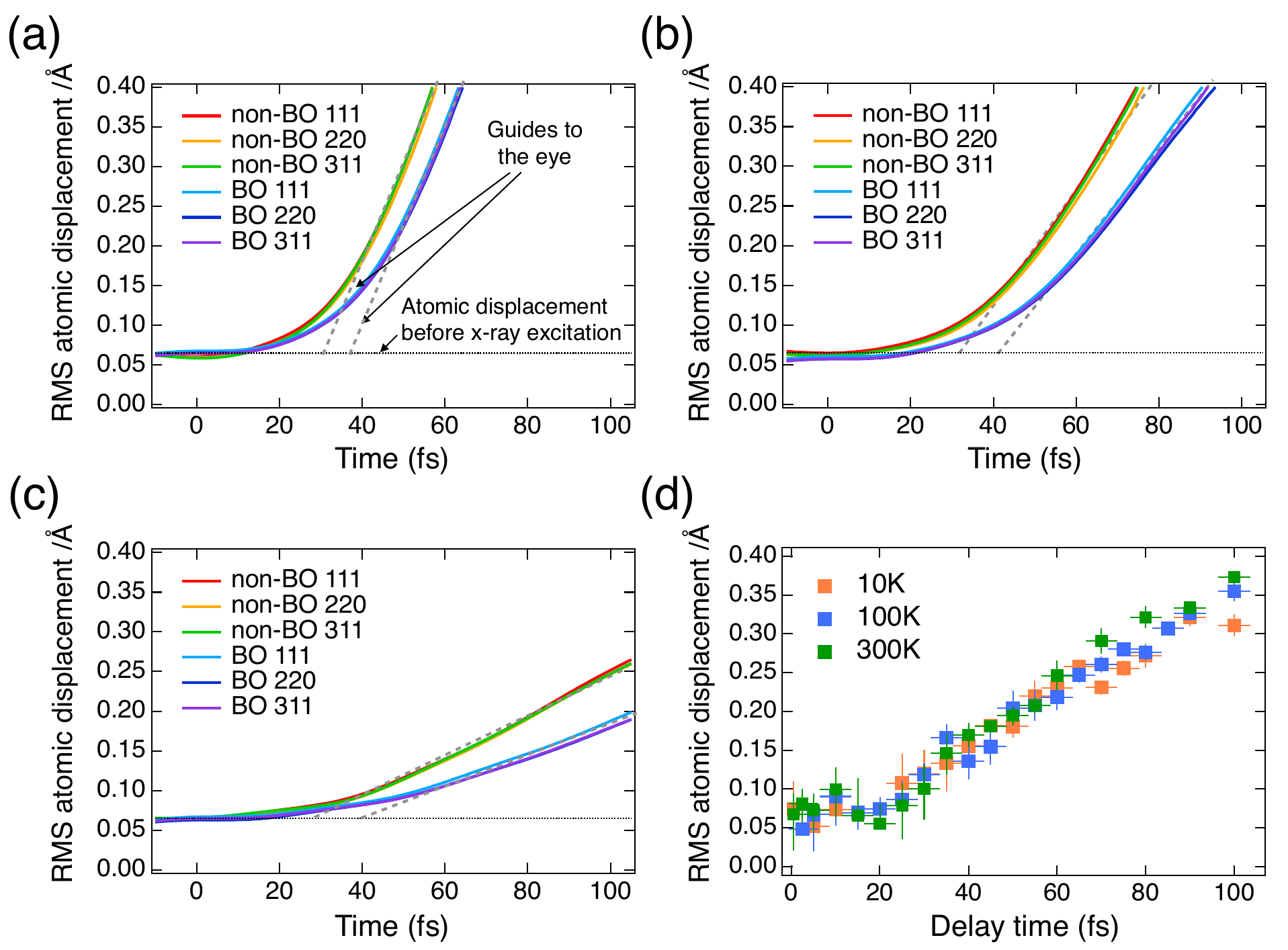}
\caption{(a-c) Root-mean-square atomic displacements perpendicular to the (111), (220) and (311) planes of silicon at 300 K predicted by non-BO and BO simulations  with fluences of (a) 100\%, (b) 48\%, and (c) 16\% of the average fluence in the experiment. (d) Experimentally observed root-mean-square atomic displacement perpendicular to the (220) plane of x-ray-excited silicon.}
\end{figure}

In summary, we conducted an x-ray pump-x-ray probe experiment on Si and observed femtosecond structural changes at various initial sample temperatures. The results of the experiment revealed that the onset time and rate of atomic disordering were not affected by the initial sample temperature. This suggests that thermal atomic motion in the equilibrium state does not play a significant role in x-ray-induced ultrafast melting. By comparing observed and simulated onset time of the atomic disordering, it was found that both thermal and non-thermal mechanisms contribute to the progress of the atomic disordering. 
The discovery of the significance of non-thermal and thermal effects in x-ray-induced ultrafast melting should encourage the development of quantitative models for intense x-ray interactions with matter. This is crucial for the planning and proper interpretation of various experiments utilizing focused XFEL pulses, such as molecular imaging \cite{Neutze2000}, protein nanocrystallography \cite{Chapman2011}, generation of warm-dense matter and plasma in the high-energy-density regime \cite{Vinko2012}, and the investigation and applications of nonlinear x-ray-matter interaction processes \cite{YonedaNature2015, TamasakuNP2014, TamasakuPRL2018, Inoue2021_2}.

\acknowledgements{The work was supported by the Japan Society for the Promotion of Science (JSPS) KAKENHI Grants (19K20604, 19KK0132, 20H04656, 21H05235, 22H03877). K. J. K. thanks the Polish National Agency for Academic Exchange for funding in the frame of the Bekker programme (PPN/BEK/2020/1/00184). }

\bibstyle{natbib}
\bibliography{Ref}

\end{document}